\documentclass[aps,pre,twocolumn,showpacs,superscriptaddress]{revtex4}
\usepackage{amssymb,amsmath,amsthm}
\usepackage{graphicx,color,epsfig}
\begin{document}

\title{Network topology and collapse of collective stable chaos}
\author{J. Gonz\'alez-Est\'evez} 
\affiliation{Laboratorio de Investigaci\'on en F\'isica Aplicada y Computacional, \\Universidad Nacional Experimental del T\'achira, 
San Crist\'obal, Venezuela.}
\author{M. G. Cosenza}
\affiliation{Centro de F\'isica Fundamental, Universidad de Los Andes, M\'erida, Venezuela.}

\begin{abstract}
Collective stable chaos consists of the persistence of disordered patterns in dynamical spatiotemporal systems possessing a negative maximum Lyapunov exponent. We analyze the role of the topology of connectivity on the emergence and collapse of collective stable chaos in  systems of coupled maps defined on a small-world networks. As local dynamics we employ a map that exhibits a period-three superstable orbit. The network is characterized by a rewiring probability $p$. We find that collective chaos is inhibited on some ranges of values of the probability $p$; instead, in these regions the system reaches a synchronized state equal to the period-three orbit of the local dynamics. Our results show that the presence of long-range interactions can induce the collapse of collective stable chaos in spatiotemporal systems.
\end{abstract}

\maketitle

\section{Introduction}
The study of dynamical processes on nonuniform or complex
networks has attracted much attention (Motter et. al., 2006; Newman, Barab\'asi and  Watts, 2006). 
In particular, the influence of the network topology have been investigated on phenomena such as synchronization (Barahona and Pecora, 2002), pattern formation (Cosenza and Tucci, 2001), spatiotemporal intermittency (Cosenza and Tucci, 2002), phase-ordering and critical phenomena (Dorogovtsev, Goltsev and Mendes, 2008), among others.

In this context, there has been interest in the research of the phenomenon of nontrivial collective behavior in dynamical networks.  This phenomenon arises when dissimilar forms of temporal evolution of macroscopic and microscopic quantities coexist in a system (Manrubia, Mikhailov and Zanette, 2004). Two different manifestations of nontrivial collective behavior have been investigated. On one hand, there is emergence of order in the time evolution of the dynamics of macroscopic quantities of a system of chaotic elements. For example, the spatial average of the states of the elements may be periodic in time, while the evolution of the individual dynamics of these elements is chaotic and desynchronized. This kind of collective behavior has been widely studied (Kaneko, 1990a; Kaneko, 1994; Chate and Manneville, 1992; Cosenza and Gonz\'alez, 1998). The counterpart of this phenomenon has also been observed: the persistence of collective chaos or temporal disorder at the macroscopic level in systems of interacting elements whose individual dynamics is periodic. This nontrivial behavior has been called {\it collective chaos} (Politi, Livi, Oppo and Kapral, 1993; Kapral, Livi, Oppo and Politi, 1994;
Kapral, Livi, Oppo and Politi, 1997) and is one of the least understood emergent phenomena in complex systems.

Stable collective chaos consists of an irregular behavior that can not be described by the presence of repellers in the phase space of the system, producing, as a consequence, the divergence of nearby trajectories. Moreover, in such systems, the duration time of the transient regime can scale exponentially with respect to the size system, and the final stable attractor can not be reached in the practical sense for large enough systems (Crutchfield and Kaneko, 1988; Kaneko, 1990b; Politi et al., 1993;
Cecconi, Livi and Politi, 1998; Bagnoli and Cecconi, 2001).

This type of irregular collective behavior that occurs in systems of coupled periodic dynamic elements is different from the phenomenon of transient chaos,
which is a truly chaotic regime with a finite lifetime, and is characterized by the coexistence of stable attractors and non-attractive chaotic sets (repellers) in the phase space of a system  (Tel and Lai, 2008). In such systems, a generic initial configuration typically produces an irregular trajectory until it suddenly collapses into a non-chaotic attractor (Wackerbauer and Showalter, 2003). In the case of stable collective chaos, the transient collective behavior, which is usually considered irrelevant, is chaotic and statistically stationary. The behavior of the system in the transient regime can not be distinguished from a typical chaotic behavior. For this reason, we speak of the existence of supertransients, since the chaotic collective behavior is, even for a system of moderately small size, the only behavior observable in practice. This phenomenon was first reported  in a network of coupled chaotic maps (Crutchfield and Kaneko, 1988; Kaneko, 1990b). This system may exhibit, for a system of $128$ elements and taking into account the speed and precision of the computer, a characteristic time of the supertransitory regime of about $10^{64}$ years (Crutchfield and Kaneko, 1988). This result has a deep impact on some physical processes that are not yet fully understood, such as turbulence in fluids or the observation of aperiodic behaviors in complex systems. Such processes could correspond, from a strictly mathematical point of view, to a transient state. In practice, we may never reach the final regular behavior. Thus, the truly stationary, observable behavior will be the supertransitory regime of collective chaos (Bagnoli and Cecconi, 2001).

Transient spatiotemporal chaos has been studied in reaction-diffusion continuous systems, such as Gray-Scott's equations
(Wackerbauer and Showalter, 2003; Wackerbauer and Kobayashi, 2007; Wackerbauer, 2007; Yonker and Wackerbauer, 2006), where it has been found that the spatial boundary conditions can induce the collapse of transient chaos towards a fixed point. On the other hand, long transients appear in networks of model neurons when the number of connetions per neuron is small
(Zillmer, Brunel and D. Hansel, 2009).

In this article we investigate the influence of the topology of the connectivity
in the process of collapse of collective chaos in networks of interacting dynamical elements.
Recently, intensive research has been performed
on the theory and applications of small-world networks (Newman, Barabasi and Watts, 2006). They constitute a class of networks
with a high degree of local clustering and a small
characteristic length between any two elements (Watts
and Strogatz, 1998). It has been
shown that small-world networks describe many natural and
artificial networks. By varying a parameter, small-world
networks can be continuously tuned between ordered, deterministic
lattices and completely random networks. 
In this paper we consider a coupled map model defined on a small-world network that exhibits the phenomenon of collective chaos and show its collapse as a result of the
variation in the connection topology of the system. Because of their discrete spatial nature, coupled map systems are especially appropriate for investigating dynamical processes occurring in heterogeneous media or in complex networks.

\section{The model.}

We use the algorithm of construction of small-world networks originally proposed by Watts and Strogatz (Watts
and Strogatz, 1998). We start from a ring with $N$ nodes, where each node is
connected to its $k$ nearest neighbors, $k$ being an even number.
Then each connection is rewired at random with probability
$p$ to any other node in the network, to avoid self-connections.
After the rewiring process, the number of elements coupled
to each site -- which we call neighbors of that site -- may vary,
but the total number of links in the network is constant and
equal to $Nk/2$. It is assumed that all links are bidirectional.
The condition $\log N \ll k \ll N $ is employed in the algorithm to ensure that no node is isolated after the rewiring process, which results in a connected graph.
For $p=0$, the network is completely regular, while for $p=1$, the resulting network is completely random. With this algorithm, a small-world network is formed for values of the probability in the range $p \in (0.001, 0.2)$.

The state of each node in the network can be characterized by a continuous variable, which evolves according to a deterministic rule depending on its own state and the states of neighboring nodes. We define a diffusively coupled map system on this network as
\begin{equation}
 x_{t+1}^i= f(x_t^i) + \frac{\epsilon}{k^{i}}\sum_{j \in \nu^{i}} \left[ f(x_t^j)-f(x_t^i) \right] \, ,
\label{eq:cml}
\end{equation}
where $x^{i}_{t}$ is the state of $i$-th map for the discrete time $t$, with $i = 1,2, \ldots, N$; $k^{i}$ is the number of elements connected to the element $i$; $\nu^{i}$ is the set of neighbors of $i$; $f(x)$ is a function that describes the local dynamics; and $\epsilon$ is the coupling parameter.

A map with minimal ingredients that presents collective chaos is (Politi et al., 1993) 
\begin{equation}
f(x) = 
\left\{
\begin{array}{ll}
bx,  & 0 < x < 1/b \\
a + c\left( x - 1/b \right),  &  1/b \leq x < 1 
\end{array}\right. ,
\label{eq:plok}
\end{equation}
where $x \in [0,1]$, and the parameters of the map $a$, $b$, and $c$ are chosen such that the local dynamics converge towards a period three superstable orbit, corresponding to the points $x_1 = a  \rightarrow x_2 = ab \rightarrow x_3 = ab^2$. A superstable orbit has a Lyapunov exponent, $\lambda$, with value $\lambda \rightarrow - \infty$. Here we shall set $a = 0.1$, $b = 2.5$, and $c = 0$. 

Note that the stability of the period-three orbit of the local map  Eq.~(\ref{eq:plok}) implies the stability of the system given Eq.~(\ref{eq:cml}), whose maximum Lyapunov exponent turns out be negative for any value of $\epsilon$. Thus, the long-term evolution of the coupled system is confined to this periodic attractor.

The collective behavior of the network can be characterized by the instantaneous mean field of the system, defined as
\begin{equation}
\label{mean}
H_t=\frac{1}{N} \sum^N_{j=1} f(x^j_t).
\end{equation}

Similarly, the state of synchronization of the elements in the system Eq.~(\ref{eq:cml}) can be characterized by the asymptotic time average $\langle \sigma \rangle$ of the instantaneous standard deviations $\sigma_t$ of the distribution of variables $x^i_t$, defined as
\begin{equation}
\sigma_t=\left[ \frac{1}{N} \sum_{i=1}^N \left( f(x^i_t) \right)^2 - H_t^2 \right]^{1/2}.
\label{eq:disper}
\end{equation}

A synchronized state corresponds to $\langle \sigma \rangle = 0$.  In our case, the system Eq.~(\ref{eq:cml}) can be synchronized in the period-three superstable orbit of the local map for certain values of parameters.

In all calculations shown below, we use networks with size $N = 10^{4}$, and assign random initial conditions $x_0^i \in [0,1]$ in the system Eq.~(\ref{eq:cml}). 

\begin{figure}[h]
\centerline{\includegraphics[width=8.5cm]{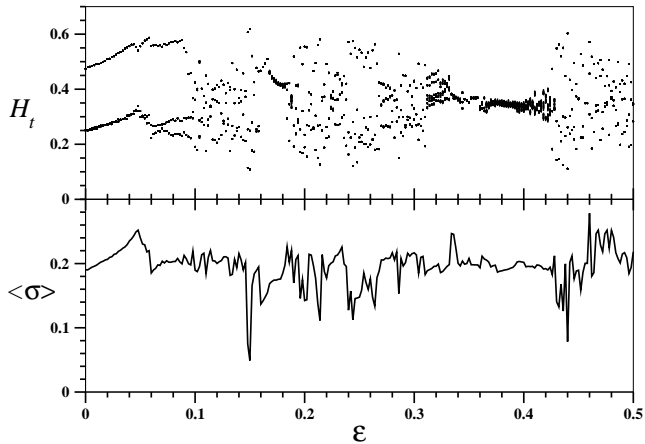}}
\caption{\small Bifurcation diagrams of the mean field $H_t$ (top) and the quantity $\langle \sigma \rangle$ (bottom) as functions of $\epsilon$, for the system Eq.~(\ref{eq:cml}) with $k = 2$ and $p = 0$ (a one-dimensional lattice), size $N = 10^{4}$. For each value of $\epsilon$,  $10^{3}$ values of $H_t$ are plotted, after discarding $10^{4}$ iterations. The quantity $\langle \sigma \rangle$ was calculated as the average of $10^{3}$ values of $\sigma_{t}$, after discarding $10^{4}$ iterations.}
\label{fig:fig1}
\end{figure}

Figure~\ref{fig:fig1} shows the bifurcation diagram of the mean field, $H_t$, and the average standard deviation $\langle \sigma \rangle$ as functions of coupling parameter $\epsilon$ for a network with $k=2$ and $p=0$, which corresponds to a one-dimensional regular lattice.

In Figure~\ref{fig:fig1} we see that the macroscopic variable $H_t$ displays chaotic behavior and chaotic bands in a wide range of the parameter $\epsilon$, in spite of the existence of period-three superstable orbit in the local maps. The global attractor of the system must be a synchronized orbit equal to the superstable period-three of the local maps. However, we observe that the quantity $\langle \sigma \rangle \neq 0$, indicating that the system does not reach a synchronized state in its long-term evolution. The convergence time to get the synchronized periodic asymptotic state is extremely large. Figure~\ref{fig:fig1} shows evidence of the phenomenon of collective stable chaos occurring in this system.

\begin{figure}[h]
\centerline{\includegraphics[width=8.5cm]{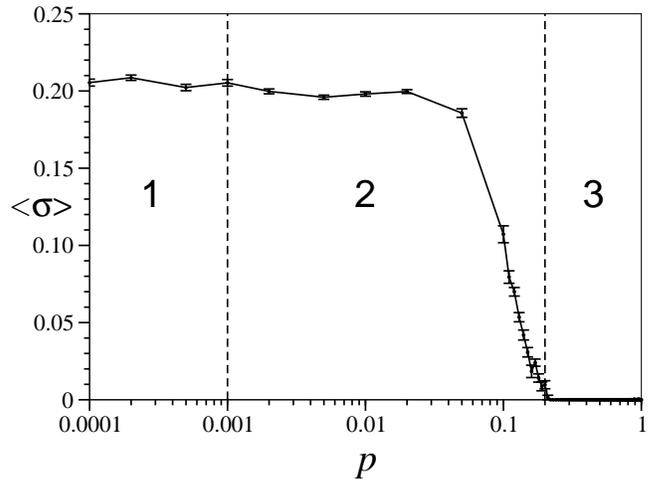}}
\caption{\small The quantity $\langle \sigma \rangle$ as a function of the reconnection probability $p$, for fixed $\epsilon = 0.384$, $k=10$, and $N=10^{4}$. The quantity $\langle \sigma \rangle$ is the average of $10^{2}$ values of $\sigma_{t}$, after discarding $3 \times 10^{4}$ iterations. 
Error bars correspond to the standard deviations of $10$ realizations of initial conditions for each value of $p$. The numbers indicate the regions of the parameter $p$ corresponding to (1) regular network, (2) small-world network, and (3) random network.}
\label{fig:fig2}
\end{figure}

To study the effects of long-range interactions in the phenomenon of stable chaos in the system Eq.~ (\ref{eq:cml}), we calculate the quantity $\langle \sigma \rangle$ as a function of the rewiring probability $p$, as shown in Figure~\ref{fig:fig2}. There is a critical value of the probability $p$ above which the system reaches a synchronized state, characterized by $\langle \sigma \rangle \rightarrow 0$. This synchronized state corresponds to the superstable orbit of superstable period-three of the local map Eq.~(\ref{eq:plok}). Thus, the presence of long-range interactions associated to an increase of $p$ contributes to the collapse of collective stable chaos and to the emergence of synchronization in this system.

The transient nature of collective chaos observed in the system  Eq.~(\ref{eq:cml}) implies that the chaotic behavior observed for certain ranges of $p$ in Fig.~\ref{fig:fig2} will eventually converge towards the stable period-three attractor of the local map. However, the regime of collective chaos is, indeed, supertransient for those parameter values. We have verified that, even for a small system consisting of $N=10$ maps, it requires about $10^{10}$ iterations to reach the synchronized period-three state. Thus, in practice, collective stable chaos becomes the actual observable macroscopic state for those values of parameters.

\section{Conclusions}
The existence of a supertransitory regime such as collective stable chaos can not be distinguished from a typical stable chaotic behavior in a system. For this reason the supertransitory states are, even for a moderately small system, the only real observable behavior.

Figure \ref{fig:fig2} shows that the system can reach the synchronized state of period three by changing the topology of the network. The presence of long-range interactions between the nodes, favored by increasing the rewiring probability $p$, facilitates the transmission of information and the possibility of synchronization in small-world and random networks, and therefore contributes to the collapse of collective stable chaos.

Our results indicate that the persistence of collective chaos in spatiotemporal systems depends on topological properties of the spatial substrate and, therefore, are not robust properties. The results also suggest that non-trivial collective behavior observed in many complex systems can be modulated or controlled through changes in their geometric properties.

\section*{Acknowledgments}
This work was supported by Consejo de Desarrollo, Cient\'ifico, Human\'istico y Tecnol\'ogico, Universidad de Los Andes, through project C-1692-10-05-B. J. G.-E. acknowledges support from Decanato de Investigaci\'on of the Universidad Nacional Experimental del T\'achira, through project 04-013-2009.

\section*{References} 
 \begin{description}
\item Bagnoli, F. and Cecconi, F. 2001. Synchronization of non-chaotic dynamical systems, Phys.
Lett A \textbf{282}: 9-17.
\item Barahona, M. and Pecora, L. M. 2002.  Synchronization in small-world systems,  Phys. Rev. Lett. \textbf{89}: 054101-4.
\item Cecconi, F., Livi, R. and Politi, A. 1998. Fuzzy transition region in a one-dimensional coupled stable-map lattice, Phys. Rev. E \textbf{57}: 2703-2712.
\item Chate, H. and Manneville, P. 1992. Emergence of effective low-dimensional dynamics in the
macroscopic behaviour of coupled map lattices, Europhys. Lett. \textbf{17}: 291-296.
\item Cosenza, M. G. and Gonz\'alez, J. 1998. Synchronization and collective behavior in globally
coupled logarithmic maps, Prog. Theor. Phys. \textbf{100}: 21-38.
\item Cosenza, M. G. and Tucci, K. 2001. Pattern formation on trees,  Phys. Rev. E \textbf{64}: 026208.
\item Cosenza, M. G. and Tucci, K. 2001. Turbulence in small-world networks, Phys. Rev. E \textbf{65}: 036223. 
\item Crutchfield, J. P. and Kaneko, K. 1988. Are attractors relevant to turbulence?, Phys. Rev. Lett.
\textbf{60}: 2715-2718.
\item Kaneko, K. 1990a. Globally coupled chaos violates law of large numbers, Phys. Rev. Lett.
\textbf{65}: 1391-1394.
\item Kaneko, K. 1990b. Supertransients, spatiotemporal intermittency, and stability of fully developed
spatiotemporal chaos, Phys. Lett. A \textbf{149}: 105-112.
\item Kaneko, K. 1994. Relevance of clustering to biological networks, Physica D \textbf{75}: 55-73.
\item Kapral, R., Livi, R., Oppo, G.-L. and Politi, A. 1994. Dynamics of complex interfaces, Phys.
Rev. E \textbf{49}: 2009-2022.
\item Kapral, R., Livi, R., Oppo, G.-L. and Politi, A. 1997. Critical behavior of complex interfaces,
Phys. Rev. Lett. \textbf{79}: 2277-2280.
\item Manrubia, S., Mikhailov, A. and Zanette, D. 2004. Emergence of dynamical order, Vol. 2 
World Scientific Lecture Notes in Complex Systems, World Scientific, Singapore.
\item Motter, A. E., Matias, M. A. , Kurths, J. and Ott, E. 2006. Dynamics on Complex Networks and Applications, Physica D \textbf{224}: vii-viii.
\item Dorogovtsev, S. N., Goltsev, A. V. and  Mendes, J. F. 2008. Critical phenomena in complex networks,  Rev. Mod. Phys. \textbf{80}: 1275-1335. \\
\item Newman, M. E. J., Barabasi, A. L. and Watts, D. J. 2006. The structure and dynamics of
networks, Princeton University Press, Princeton, N. J.
\item Politi, A., Livi, R., Oppo, G.-L. and Kapral, R. 1993. Unpredictable behavior in stable systems,
Europhys. Lett. \textbf{22}: 571-576.
\item Tel, T and Lai,  Y. C. 2008. 
Chaotic transients in spatially extended systems,  Physics  Reports \textbf{60}: 245-275.
\item Wackerbauer, R. 2007. Master stability analysis in transient spatiotemporal chaos, Phys. Rev.
E \textbf{76}: 056207.
\item Wackerbauer, R. and Kobayashi, S. 2007. Noise can delay and advance the collapse of spatiotemporal
chaos, Phys. Rev. E 75: 066209.
\item Wackerbauer, R. and Showalter, K. 2003. Collapse of spatiotemporal chaos, Phys. Rev. Lett.
\textbf{91}: 174103.
\item Watts, D. J. and Strogatz, S. H. 1998. Collective dynamics of “small-world” networks, Nature
\textbf{393}: 440-442.
\item Yonker, S. and Wackerbauer, R. 2006. Nonlocal coupling can prevent the collapse of spatiotemporal
chaos, Phys. Rev. E \textbf{73}: 026218.
\item Zillmer, R., Brunel, N. and Hansel, D. 2009. Very long transients, irregular firing, and chaotic dynamics in networks of randomly connected inhibitory integrate-and-fire neurons,  Phys. Rev. E \textbf{79}: 031909.
\end{description}
\end{document}